\documentclass[12pt]{article}
\usepackage{amsmath}
\usepackage{graphicx}
\usepackage{times}
\usepackage[T1]{fontenc}
\usepackage[utf8]{inputenc}
\usepackage[margin=1cm]{caption}

\usepackage[square,numbers,sort]{natbib}

\author
{Shiva Gaur$^{1}$, Akash Kumar$^{2,3,4}$, 
 Himanshu Bangar$^{5}$, Utkarsh Shashank$^{2}$, \\Hukum Singh$^{1}$, Saroj P. Dash$^{5}$, Anubhav Raghav$^{1\ast}$
and Johan \AA kerman$^{2,3,4\ast}$
\\
\normalsize{$^{1}$Department of Applied Sciences, The NorthCap University,}\\
\normalsize{Gurugram, Haryana-122017, India}\\
\normalsize{$^{2}$Applied Spintronics Group, Department of Physics, University of Gothenburg, Fysikgränd 3,}\\
\normalsize{412 96, Gothenburg, Sweden}\\
\normalsize{$^{3}$Research Institute of Electrical Communication, Tohoku University, 2-1-1 Katahira,}\\
\normalsize{Aoba-ku, Sendai 980-8577 Japan}\\
\normalsize{$^{4}$Center for Science and Innovation in Spintronics, Tohoku University, 2-1-1 Katahira,}\\
\normalsize{Aoba-ku, Sendai 980-8577 Japan}\\
\normalsize{$^{5}$Department of Microtechnology and Nanoscience, Chalmers University of Technology}\\
\normalsize{412 96, Gothenburg, Sweden}\\
\\
\normalsize{$^\ast$To whom correspondence should be addressed; E-mails:} \\
\normalsize{anubhavraghav19292@gmail.com, johan.akerman@physics.gu.se}
}

\date{}

\begin{document}

\parindent 0cm
\parskip 12pt

\title{\LARGE\bfseries{Giant Spin Pumping at Polymer/Ferromagnet Interfaces for Hybrid Spintronic Devices}} 

\maketitle

\begin{abstract}
While the growing utilization of polymers in flexible electronic devices has sparked significant interest in polymer/metal interfaces, spintronic studies of such interfaces remain limited. Here, we systematically study spin pumping across a polymer/ferromagnet metal interface between hydrogen silsesquioxane (HSQ) oligomer layers ($t_\mathit{HSQ} = 30, 36, 48$ nm) and NiFe ($t_\mathit{NiFe} = 4, 5, 7, 10$ nm) thin films. Using ferromagnetic resonance measurements, we observe strong spin pumping (large linewidth broadening) and a giant spin mixing conductance, reaching 19.8~${\rm nm^{-2}}$ for HSQ = 48 nm, \emph{i.e.}~comparable to that of heavy metals. Our results suggest efficient spin transfer across the HSQ/NiFe interface, possibly originating from a combination of spin and orbital pumping, and provide valuable insights for designing self-powered and flexible spintronic devices utilizing polymers in combination with ferromagnetic materials.
\end{abstract}

\section{Introduction}

Over the past decades, spintronics has emerged as a promising technology to complement or even replace charge-based microelectronics, owing to its advantages such as non-volatility, low power consumption, high speed, and radiation hardness~\cite{dieny2020opportunities}. A fundamental aspect of spintronic devices is charge-to-spin conversion, typically achieved via the spin Hall effect (SHE) in heavy metal/ferromagnet (HM/FM) bilayers~\cite{hirsch1999spin, sinova2015spin}. Conversely, spin pumping—the reciprocal effect of the SHE—transfers pure spin current from a precessing FM layer to a non-magnetic HM layer~\cite{tserkovnyak2002enhanced, tserkovnyak2005nonlocal}. This mechanism has been extensively employed for THz generation in FM/HM heterostructures~\cite{seifert2016efficient} and for investigating spin transport phenomena at FM/non-magnetic interfaces, including those involving non-magnetic metals, two-dimensional materials, and topological insulators~\cite{kajiwara2010transmission, noel2020non, bansal2019extrinsic, bangar2022large}. However, significant spin pumping has primarily been observed in heavy metals with strong spin-orbit coupling, such as Pt, W, and Ta~\cite{ando2008angular, nakayama2012geometry, jhajhria2019dependence}.

Recently, spin pumping has been extended to polymer-based systems, particularly organic materials, enabling the assessment of spin-orbit coupling and spin injection efficiency in these materials~\cite{devkota2016organic, vetter2020tuning, tahir2025verdazyl, tahir2025spintronic}. Polymers are particularly attractive for spintronic applications due to their lightweight, flexibility, corrosion resistance, and environmental friendliness~\cite{guo2019recent}. Significant research efforts have focused on interfacing polymers with ferromagnetic materials~\cite{mostufa2024spintronic, gupta2023chemical,serrano2019two} to develop flexible spintronic devices~\cite{muscas2021ultralow} and explore the unique properties emerging from these hybrid systems~\cite{li2022sunlight, guo2024emerging}. These advancements could lead to low-cost, potentially biocompatible spintronic devices~\cite{mostufa2024spintronic} and enable integration with other technologies, such as solar cells~\cite{li2022sunlight}, wearable healthcare applications, and polymer-based magneto-ionic systems~\cite{pachat2022magneto}.

Despite these advancements, a fundamental challenge remains: the large disparity in electrical conductivity when transferring spin currents from metallic FM layers to semiconducting, insulating, or polymer materials~\cite{schmidt2000fundamental, rojas2014spin}. This has led to skepticism regarding reports of significant spin pumping in such heterostructures~\cite{lu2024semiconducting}. However, recent experimental and theoretical studies suggest that orbital torque and orbital pumping—counterparts to spin-orbit torque and spin pumping—could play a crucial role~\cite{hayashi2024observation}. Unlike the SHE, the orbital Hall effect (OHE) can be substantially stronger in lighter elements, including Si and 3d transition metals~\cite{sala2022giant}. Moreover, the OHE is less affected by the large electrical conductivity mismatch at interfaces, making it a promising mechanism for overcoming spin injection limitations in semiconductors, insulators, and polymers. Building on previous observations of local orbital angular momentum in p-doped Si~\cite{bernevig2005orbitronics}, as well as the inverse SHE and large anomalous Hall effect in Si~\cite{manyala2004large, ando2012observation}, silicon has re-emerged as a promising candidate for OHE and spintronics~\cite{baek2021negative, matsumoto2025observation}. While organic polymers have been explored in spintronics, silicon-based inorganic polymers remain largely unexplored, despite their abundance and unique properties~\cite{jin2022lambda}.

In this work, we investigate the spin pumping behavior at the interface of a ferromagnetic metal (NiFe) and a silicon-based cross-linking polymer, hydrogen silsesquioxane (HSQ, $\rm H_{\rm 8}\rm Si_{\rm 8}\rm O_{\rm 12}$). HSQ is a highly ordered oligomer widely used as a negative resist in photolithography and electron beam lithography~\cite{peuker2002hydrogen, jin2022lambda}, as well as a planarizing interlayer dielectric film in integrated circuits~\cite{yang2002structures}. By analyzing ferromagnetic resonance (FMR) linewidth broadening, we observed significant spin pumping from the NiFe layer into the non-magnetic HSQ polymer, as evidenced by a comparison with bare NiFe control samples. Systematic thickness-dependent studies of the ferromagnetic layer enabled us to determine the spin mixing conductance for various HSQ thicknesses, with the highest value, $19.8~{\rm nm^{-2}}$, observed for an HSQ thickness of 48 nm. This value exceeds those reported for other organic polymers~\cite{manoj2021giant, wittmann2020tuning} and metal-free conjugated polymers~\cite{vetter2020tuning}, and is comparable to non-conjugated polymers~\cite{tahir2025spintronic, tahir2025verdazyl} and heavy metals with strong spin-orbit coupling, such as Ta, Pt, W, and Pd~\cite{kumar2018large, ando2008angular, ando2009electric, nakayama2012geometry}.

\begin{figure*}
 \centering
 \includegraphics[width=\linewidth]{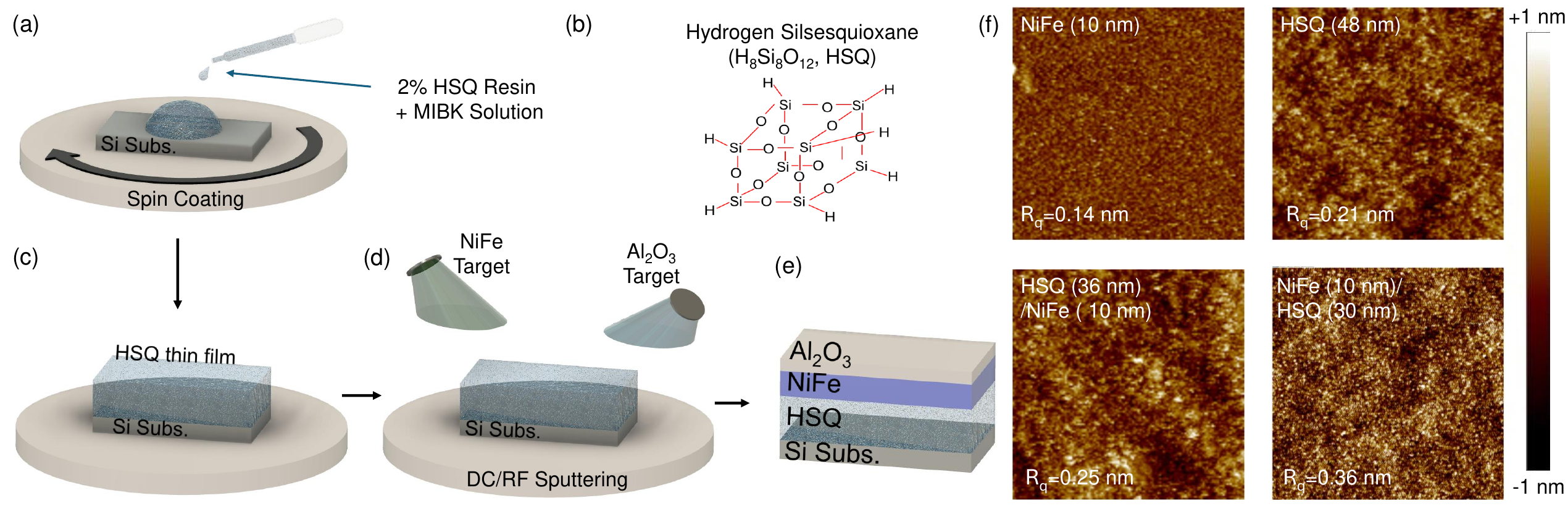}
\caption{(a-e) Complete process of sample preparation; (a) Schematic depiction of preparing thin film of HSQ polymer using spin coating using HSQ resin solution, (b) the atomic structure of HSQ oligomer, (c) wet thin film of HSQ after spin coating (d) sputtering configuration for metal deposition (e) final thin film stack structures. (f) respective characterization with AFM over a  $1\times 1~\mu m^2$ scanning area of a bare NiFe (10 nm) thin film sample (without HSQ), HSQ thin film without ferromagnet NiFe, HSQ/NiFe stacks and reversed stack. }
\end{figure*}

\section {Results}

\subsection{Sample structure and characterization}

Bilayer thin films of HSQ/NiFe were fabricated using a combination of spin coating and DC/RF magnetron sputtering, as illustrated in Fig.~1(a–e). HSQ is a low-dielectric inorganic oligomer with a cage-like structure composed of silicon and oxygen atoms with dangling hydrogen atoms (see Fig.~1b)~\cite{ro2011silsesquioxanes, jin2022lambda}. In solution, the oligomer forms amorphous SiO:H layers with excellent planarization properties~\cite{yang2002structures}. Further details on sample preparation are provided in the Methods section.

The surface morphology of the HSQ thin films was characterized using atomic force microscopy (AFM) in tapping mode for four sample configurations: (i) pristine HSQ, (ii) HSQ after NiFe deposition, (iii) reference NiFe, and (iv) HSQ deposited on top of NiFe. The AFM topographic images in Fig.~1(f) reveal a homogeneous surface morphology across all samples, with no observable pinholes or significant surface irregularities. Quantitative analysis indicates that the average surface roughness varies from 0.14 nm for bare NiFe (10 nm) and 0.21 nm for HSQ (48 nm) to a maximum of 0.36 nm for the bilayer configuration of HSQ (30 nm)/NiFe (10 nm). These exceptionally low roughness values confirm that NiFe deposition has a negligible impact on the underlying HSQ layers~\cite{tahir2025verdazyl}. Furthermore, the quality of the NiFe thin films deposited on HSQ remains largely preserved, underscoring the compatibility of the deposition process with spin-coated HSQ thin films.

To assess electrical properties, we measured the resistivity of HSQ/NiFe bilayers with NiFe (10 nm) for all HSQ thicknesses using the four-point probe technique (data provided in the Supplementary Information). The measured values are consistent with previously reported resistivities for NiFe thin films~\cite{mazraati2018improving}. The reference sample (without HSQ) exhibited a lower resistivity compared to the HSQ-coated samples, which is attributed to the shunting effect in bare NiFe samples on semiconducting Si substrates~\cite{behera2024ultra}. This confirms that HSQ acts as an insulating layer.

\subsection{Spin pumping in HSQ/NiFe heterostructure: HSQ thickness dependence}

To evaluate spin pumping at the HSQ/NiFe interface, FMR spectra were recorded for bilayer HSQ/NiFe films with varying HSQ thicknesses. The results were compared with reference NiFe films without HSQ. FMR measurements were performed using a NanOsc Phase-40 FMR instrument, and a schematic of the co-planar waveguide (CPW) with the static ($H$) and RF magnetic ($h_{\rm rf}$) field directions is shown in Fig.~2a. Further details of the experimental setup can be found in~\cite{bainsla2022ultrathin,gaur2023spin} and in the 'Methods' section. Fig.~2b compares the FMR spectra (linewidth and resonance field) of the HSQ (36 nm)/NiFe (4 nm) sample with the NiFe (4 nm) references sample. Fig.~2c shows the frequency-dependent FMR spectra for the HSQ (36 nm)/NiFe (4 nm) film over the 3–15 GHz frequency range. Similar measurements were conducted for HSQ layers of 30, 36, and 48 nm with a 4 nm NiFe layer, as well as for the reference NiFe samples without HSQ. The raw spectra were analyzed by fitting them to a sum of symmetric and anti-symmetric Lorentzian derivative functions, following the method described in ~\cite{kumar2018large,bainsla2022ultrathin}: 

\begin{figure*}
 \centering
 \includegraphics[width=\linewidth]{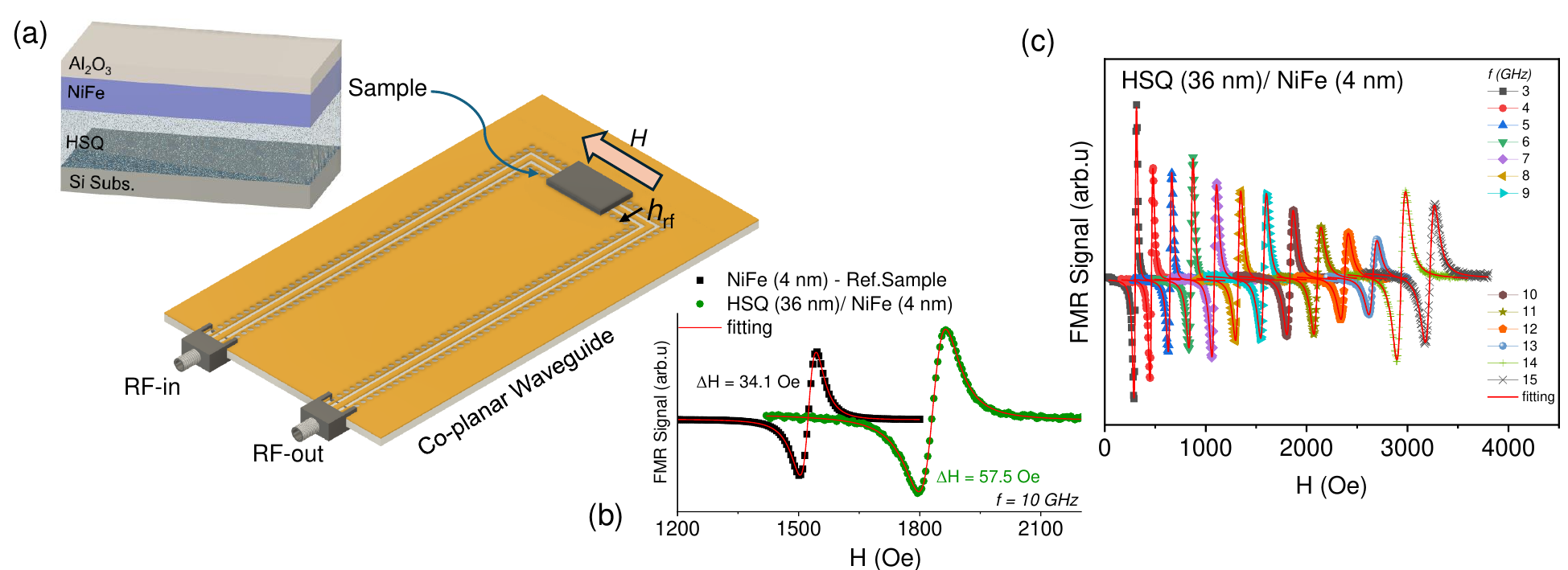}
\caption{(a) Schematic of the ferromagnetic resonance (FMR) measurement geometry. (b) Comparison of the reference sample of NiFe (4 nm) and HSQ (36 nm)/NiFe(4 nm). (c) FMR signal for wide frequency range of 3-15 GHz for HSQ (36 nm)/NiFe (4 nm) thin film sample.}
\end{figure*}

\begin{equation}\frac{d \chi_{\rm FMR}}{dH}\propto - \frac{2V_{\rm S}\Delta(H-H_{\rm R})}{(\Delta H^2+(H-H_{\rm R})^2)^2}\\ +\frac{V_{\rm AS} (\Delta H^2 -(H-H_{\rm R})^2)}{(\Delta H^2+(H-H_{\rm R})^2)^2}
\end{equation}
\label{eq: FMR}
Here, $V_{\rm S}$ and $V_{\rm AS}$ represent the symmetric and anti-symmetric components. The applied field, linewidth, and resonance fields are represented by \textit{H}, $\Delta H$ and $H_R$ respectively.

The $\Delta H$ and $H_{\rm R} $ are extracted for all samples at different thicknesses for further calculations of the effective Gilbert damping ($\alpha_{\rm eff}$) and effective saturation magnetization ($M_{\rm eff}$). The frequency-dependent $\Delta H$ and $H_R$ curves for NiFe (4 nm) samples with varying HSQ thicknesses are shown in Fig.~3a and 3b. $\alpha_{\rm eff}$ is extracted from the slope of linear fits of $\Delta H$ \emph{vs.}~frequency (Fig.~3a) using the following equation: 
\begin{equation}
 \Delta H= \Delta H_0+\frac{2\pi \alpha_{\rm eff}}{\gamma}f
\end{equation}
Here, $\Delta H_0$ indicates the inhomogeneous broadening and $\gamma= \rm 29.6$ GHz/T the gyromagnetic ratio. 
Fits to Eq.~2 are shown as solid lines in Fig.~3a. $\Delta H_{0}$ is found to be $<$12 Oe for all samples, confirming the high quality of NiFe thin films on top of HSQ. The increase in $\Delta H$ and its slope \emph{vs.}~frequency in the HSQ samples, compared to the reference samples, clearly indicates large spin pumping into the adjacent HSQ layer. We observed an $81\%$ $\alpha_{\rm eff}$ enhancement for HSQ(48 nm)/NiFe (4 nm) ($\alpha_{\rm eff}$=0.022) as compared to the reference sample ($\alpha_{\rm eff}$=0.012), highlighting the substantial spin pumping across the HSQ/NiFe interface.

Further, to calculate $M_{\rm eff}$, $f$ \emph{vs.}~$H_R$ (Fig.~3b) is fitted using the Kittel formula~\cite{kittel1948theory}:
\begin{equation}
 f= \frac{\gamma}{2\pi} \sqrt{(H_R+H_K)(H_R+H_K+4\pi M_{\rm eff})},
\end{equation}
where $H_K$ represents the in-plane magnetic anisotropy constant. The frequency \emph{vs.} $H_R$ data for all HSQ thicknesses are found to be consistent with similar effective magnetization values, again showing the uniformity of NiFe thin films grown on HSQ.

\begin{figure}
    \centering
    \includegraphics[width=0.4\linewidth]{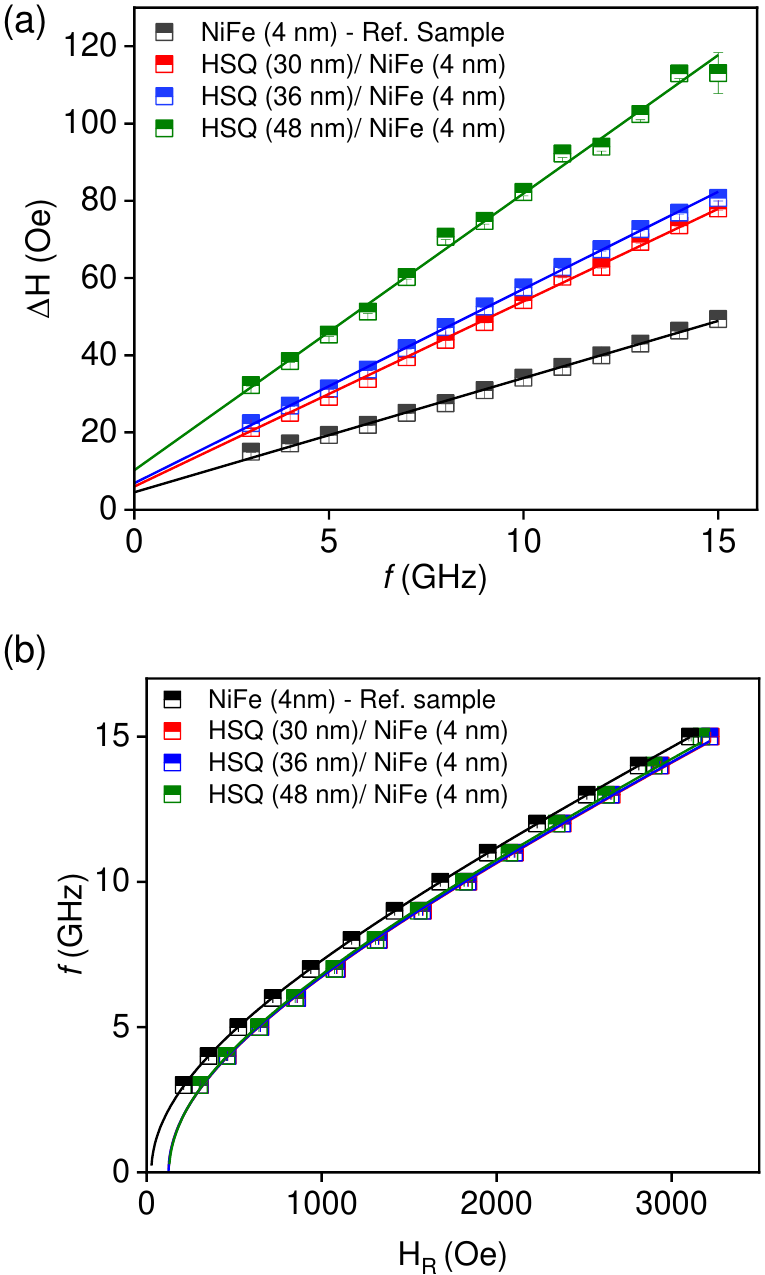}
    \caption{(a) Linewidth \emph{vs.} frequency and (b) frequency \emph{vs.}~resonance field, for reference sample and other HSQ thicknesses with constant NiFe = 4 nm.}
\end{figure}

\subsection{Ferromagnet thickness dependence and spin mixing conductance}

Fig.~4a presents the effective damping coefficient ($\alpha_{\rm eff}$) as a function of the inverse NiFe thickness for all HSQ thicknesses ($t_{\rm HSQ}$ = 0, 30, 36, and 48 nm). While the increase in damping with HSQ thickness could arise from mechanisms such as spin memory loss~\cite{rojas2014spin, bonell2020control} and two-magnon scattering~\cite{arias1999extrinsic}, the linear dependence of linewidth on frequency (Fig.~3a) rules out significant contributions from two-magnon scattering or phonon drag~\cite{arias1999extrinsic, li2021drag}. The observed linear increase in $\alpha_{\rm eff}$ with the inverse ferromagnet thickness suggests minimal spin memory loss at the interface, confirming HSQ as an efficient spin sink~\cite{tserkovnyak2002enhanced, bainsla2022ultrathin, bangar2022large, rojas2014spin}.

To quantify the spin pumping efficiency, we determine the effective spin mixing conductance (${g^{\uparrow\downarrow}_ {\rm eff}}$) by fitting $\alpha_{\rm eff}$ \emph{vs.} $1/t_{\rm NiFe}$ using the following equation~\cite{kumar2018large}:

\begin{equation}\label{eq:spi nmixing}
\alpha_{\rm eff} = \alpha_{\rm 0} + \gamma\hbar \frac{{g^{\uparrow\downarrow} _{\rm eff}}}{4\pi M_s}\frac{1}{t_{\rm NiFe}}
\end{equation}

Here, $\alpha_{\rm 0}$ represents the intrinsic Gilbert damping (bulk) of NiFe, and $\hbar$ is the reduced Planck’s constant. The saturation magnetization (${4\pi M_s}$), obtained from a linear fit of $M_{\rm eff}$ \emph{vs.}~$1/t_{\rm NiFe}$ [see Fig.~4(b)], is found to be approximately 1.05 T for all sample series [inset of Fig.~4(b)]. Fig.~4c summarizes $g^{\uparrow\downarrow}_{\rm eff}$ for different HSQ thicknesses, with a dotted black line as a visual guide. The extracted values are $g^{\uparrow\downarrow}_{\rm eff} = 7.9 \pm 0.8$, $10.5 \pm 1.4$, $15.5 \pm 1.6$, and $19.8 \pm 2.4$ nm$^{-2}$ for the reference, 30 nm, 36 nm, and 48 nm HSQ samples, respectively.

The lower $g^{\uparrow\downarrow}_{\rm eff}$ in the reference sample is attributed to minimal spin pumping into the substrate and the AlO$_{\rm x}$ capping layer. The intrinsic damping values ($\alpha_{\rm 0}$) for all samples range from 0.0030 to 0.0045, aligning with previously reported NiFe values~\cite{inaba2006damping}. Interestingly, samples with HSQ exhibit \emph{lower} $\alpha_{\rm 0}$ than the reference NiFe, potentially due to the formation of a thin silicide layer at the Si/NiFe interface~\cite{schulz2021increase}.

To further analyze spin transport, we calculated the spin current density ($J^{\rm eff}_{\rm s}$) using the following equation~\cite{mosendz2010detection}:

\begin{equation}\label{Js}
J^{\rm eff}_{\rm s}=\frac{2e}{\hbar}\times\frac{\hbar\omega}{4\pi} \mathsf{g_{\uparrow\downarrow}} \sin^2\theta_{\rm C}\times \frac{2\omega[M_{s}\gamma+\sqrt{(M_{s}\gamma)^2+(2\omega)^2}]}{(M_{s}\gamma)^2+(2\omega)^2}, 
\end{equation}

\begin{figure*}
    \centering
    \includegraphics[width=0.99\linewidth]{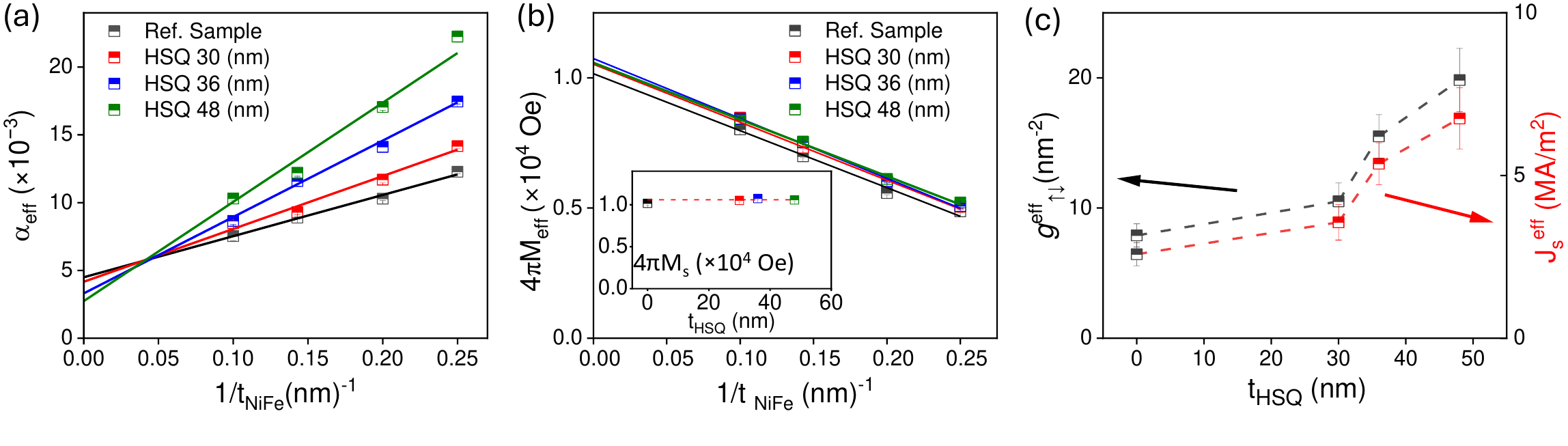}  \caption{(a) Effective Gilbert damping \emph{vs.} the inverse of NiFe thickness (1/t$_{\rm NiFe}$) for all samples. (b) Saturation Magnetization ($4\pi M_{\rm S}$) \emph{vs.} inverse thickness of NiFe, inset shows the saturation magnetization for HSQ thickness i.e the intercept of Fig. (b). (c) The resultant spin mixing conductance and spin current density parameter \emph{vs.} HSQ thickness (t$_{HSQ}$=0 nm represents the reference NiFe sample without HSQ), calculated using Eq. (4) and (5).}
\end{figure*}

Here, $e$ represents the electron charge, and $\omega$ is the angular frequency. The magnetization precession cone angle, $\theta_{\rm C}$, is given by $\theta_{\rm C} = \frac {h_{\rm rf}}{2\Delta H}$, where ${h_{\rm rf}}$ is the strength of the RF field experienced by the sample. The final term, $\frac{2\omega[M_{s}\gamma+\sqrt{(M_{s}\gamma)^2+(2\omega)^2}]}{(M_{s}\gamma)^2+(2\omega)^2}$, represents an ellipticity correction factor, which accounts for the ellipticity of the magnetization precession.

The calculated values of $J_{\rm s}^{\rm eff}$ are plotted in Fig.4c (indicated by the red visual guide) and exhibit a similar trend as $g^{\uparrow\downarrow}_{\rm eff}$. Moreover, the $J^{\rm eff}_{\rm s}$ values are comparable to those observed at FM/HM interfaces, such as Co/Pt and NiFe/Ta~\cite{kumar2019influence,kumar2018large}, confirming the generation of a significant spin current in the HSQ/NiFe heterostructures.

\subsection{Reversed sample structures: NiFe/HSQ heterostructure}

Notably, both $g^{\uparrow\downarrow}_{\rm eff}$ and $J^{\rm eff}_{\rm s}$ increase with HSQ thickness, indicating a large spin diffusion length in HSQ thin films and confirming the bulk nature of the polymer as a spin sink. However, measurements for thicker HSQ films were not feasible, as they require a more viscous HSQ resin solution in MIBK, which leads to significant surface degradation and increased interfacial roughness.

To verify that the observed effects are not caused by permanent modifications to the HSQ thin films due to high-energy bombardment during sputtering, we fabricated samples with a reversed stack order: NiFe thin films were first deposited on HiR-Si substrates via DC sputtering, followed by spin coating of HSQ thin films. The process was carried out with minimal delay to prevent NiFe oxidation.

Fig.~5 presents the effective Gilbert damping as a function of HSQ thickness, with NiFe thickness fixed at 10 nm. The continuous increase in effective Gilbert damping with increasing HSQ thickness further confirms the bulk nature of spin pumping in HSQ thin films, ruling out any structural or interfacial degradation. Additionally, this result demonstrates that the quality of the HSQ layer remains unchanged regardless of its deposition order relative to NiFe.

\subsection{Comparison with HSQ/CoFeB interface}

To determine whether spin pumping is exclusive to NiFe, we fabricated an additional sample series with CoFeB thin films: HiR-Si Subs./HSQ (30, 36, 48 nm)/CoFeB (4 nm)/Al$_2$O$_3$ (4 nm) along with a reference sample HiR-Si Subs./CoFeB (4 nm)/Al$_2$O$_3$ (4 nm). FMR measurements for varying HSQ thickness (data in supplementary information) revealed a trend similar to that observed in the HSQ/NiFe and NiFe/HSQ systems. 

Compared to the 81\% enhancement reported above for HSQ/NiFe, we observed an even higher 95\% increase in effective Gilbert damping for HSQ (48 nm)/CoFeB, compared to its reference ($\alpha_{eff}$ = 0.0094 \emph{vs.}~0.0048). 
This further supports our results and confirms the bulk nature of spin pumping in HSQ thin films regardless of the details of the ferromagnetic metal. The linewidth \emph{vs.}~frequency data and effective Gilbert damping as a function of HSQ thickness are provided in the supplementary figures.

\begin{figure}
 \centering
 \includegraphics[scale=0.43]{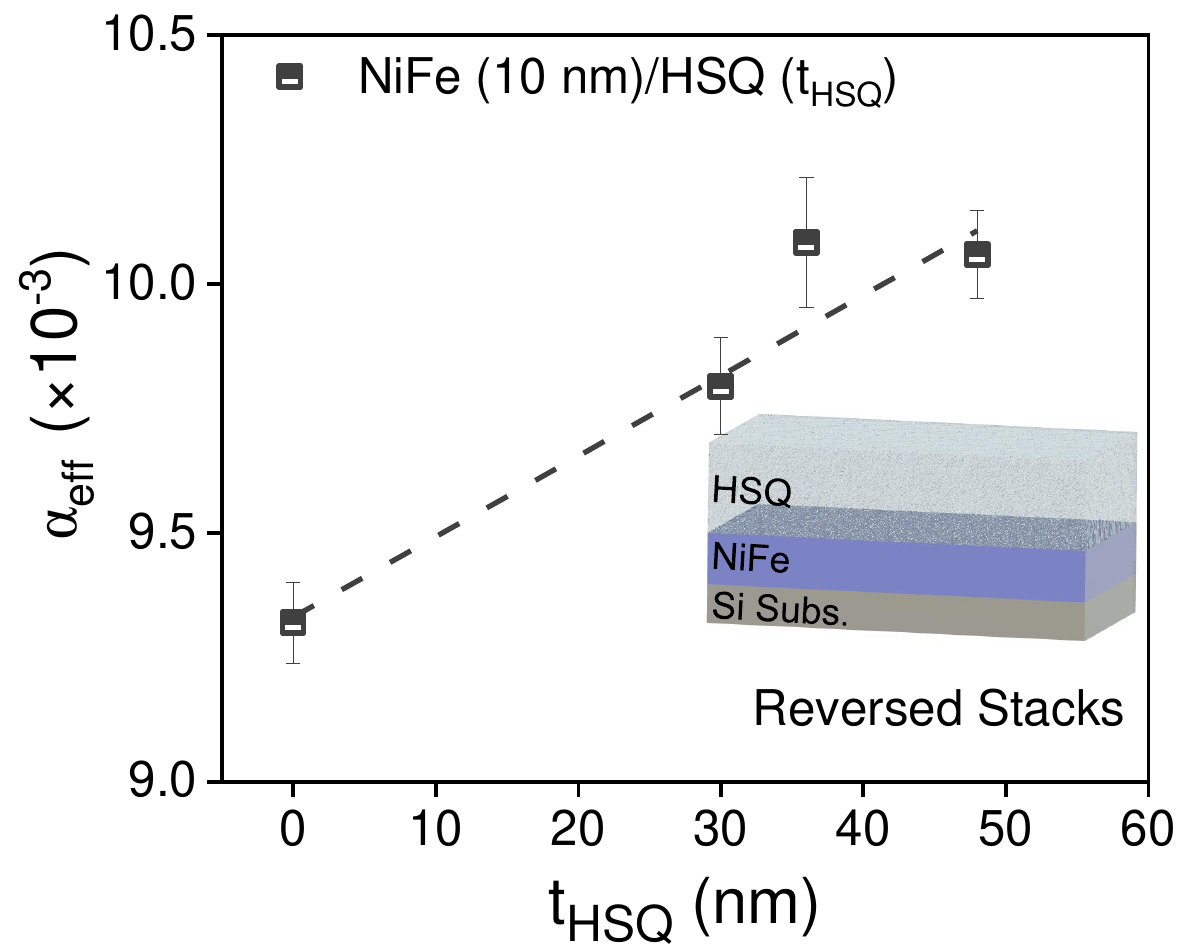}
\caption{Effective Gilbert damping \emph{vs.} HSQ thickness for the reversed stack, with fixed t$_{\rm NiFe}$=10 nm. The dotted line shows a guide to the eye. }
\end{figure}

\section{Discussion}

Earlier studies on FM/organic semiconductor polymer interfaces have also demonstrated significant spin pumping and spin mixing conductance. The organic semiconductors interfaced with NiFe thin films exhibit  ``spinterface" effects, where interaction-induced spin-dependent density of states influences spin transport~\cite{cinchetti2017activating, sanvito2010rise, pandey2023perspective}. Recent studies on conjugated polymers~\cite{manoj2021giant, gupta2023chemical, vetter2020tuning, wittmann2020tuning} further confirm strong spin pumping effects. 

Talluri et al. \cite{manoj2021giant} reported $g^{\uparrow\downarrow}_{\rm eff} = 1.54~\rm nm^{-2}$ for perylene diimide (PDI)/NiFe, while Gupta et al. found $g^{\uparrow\downarrow}_{\rm eff} \approx 1.02~\rm nm^{-2}$ for a pyrene oligomer/NiFe system. In another work, Vetter et al.~\cite{vetter2020tuning} investigated P3MT (polymethylthiopene conjugated polymer brush) and P3HT (poly3-hexylthiopene) as non-magnetic layers in NiFe-based heterostructures. 
Their results showed that the out-of-plane columnar arrangement in P3MT enhanced the damping, yielding $g^{\uparrow\downarrow}_{\rm eff} = 1.06 \times 10^{21}\rm m^{-2}$, nearly two orders of magnitude higher than conventional metallic and semiconductor systems ($10^{18}$–$10^{19}\rm m^{-2}$). For P3HT, they reported $g^{\uparrow\downarrow}_{\rm eff} = 4.22 \times 10^{20}\rm m^{-2}$. Additionally, Wittmann et al.~\cite{wittmann2020tuning} examined spin pumping in a small molecule based on $\rm DNTT-\rm dinaphtho[2,3-\textit{b}:2,3,\textit{f}] \rm thieno[3,2-\textit{b}]\rm thiophene$  with NiFe, reporting $g^{\uparrow\downarrow}_{\rm eff} = 3.35~\rm nm^{-2}$ for DNTT, $2.98~\rm nm^{-2}$ for Ph-DNTT-Ph, and $0.63~\rm nm^{-2}$ for C8-DNTT-C8, demonstrating the influence of molecular structure on spin injection efficiency.

\begin{table}[h]
    \centering
    \begin{tabular}{|l|c|l|}
        \hline
        \textbf{Materials} & \textbf{ $g^{\uparrow\downarrow}_{\rm eff}$} ($nm^{-2}$) & \textbf{Reference} \\
        \hline
        Heavy Metals & 1–30 & \cite{nakayama2012geometry,kumar2018large,jhajhria2019dependence} \\
        PTEO & 11.8 & Tahir et al.~\cite{tahir2025spintronic} \\
        PVEO & 32 & Tahir et al.~\cite{tahir2025verdazyl}\\
        DNTT & 3.35 & Wittmann et al.~\cite{wittmann2020tuning}\\
        Ph-DNTT-Ph & 2.98 & Wittmann et al.~\cite{wittmann2020tuning} \\
        C8-DNTT-C8 & 0.63 & Wittmann et al.~\cite{wittmann2020tuning} \\
        P3HT* & 0.42$\times10^3$ & Vetter et al.~\cite{vetter2020tuning} \\
        P3MT* & 1.06$\times10^3$ & Vetter et al.~\cite{vetter2020tuning}\\
        PDI & 1.54 & Talluri et al.~\cite{manoj2021giant}\\
        Pyrene & 1.02 & Gupta et al.~\cite{gupta2023chemical} \\
        HSQ & 19.8 & This Work ~\\
        \hline
    \end{tabular}
    \caption{Comparison of spin mixing conductance ($g^{\uparrow\downarrow}_{\rm eff}$) for various materials. The P3HT and P3MT represent the $g^{\uparrow\downarrow}_{\rm eff}$ at infinite polymer thicknesses~\cite{vetter2020tuning}. }
    \label{tab:spin_mixing}
\end{table}

In recent studies, Tahir et al. \cite{tahir2025spintronic,tahir2025verdazyl} investigated non-conjugated paramagnetic radical polymers, PTEO and PVEO, as efficient spin sources, reporting $g^{\uparrow\downarrow}_{\rm eff} = 11.8$ and $32.0~\rm nm^{-2}$, respectively—values significantly higher than those observed in heavy metals. In our study, HSQ (48 nm) thin films exhibit $g^{\uparrow\downarrow}_{\rm eff}$ = 19.83 ± 2.43~${\rm nm^{-2}}$, nearly an order of magnitude higher than previously reported conjugated polymers~\cite{manoj2021giant,wittmann2020tuning,gupta2023chemical} and comparable to values reported in \cite{tahir2025spintronic,tahir2025verdazyl}. 

Table 1 summaries the $g^{\uparrow\downarrow}_{\rm eff}$ values reported for various polymer/ferromagnet interfaces. Notably, these $g^{\uparrow\downarrow}_{\rm eff}$ values are comparable to those of conventional heavy metals such as Pt, W, and Ta, which are well known for their strong spin-orbit coupling and spin Hall effect~\cite{nakayama2012geometry,kumar2018large,jhajhria2019dependence}. This finding highlights the potential of polymers as efficient spin conductors. A benchmarking of $g^{\uparrow\downarrow}_{\rm eff}$ across different material classes is provided in the supplementary information.

High spin pumping and large $g^{\uparrow\downarrow}_{\rm eff}$ are typically found in metallic non-magnets interfaced with ferromagnets that have similar electrical resistivity, clean interfaces, and strong spin-orbit coupling~\cite{ando2008angular,rojas2014spin}. In contrast, insulating non-magnets present a fundamental block for spin injection from metallic ferromagnets~\cite{schmidt2000fundamental}. Since HSQ is an insulating material with weak spin-orbit coupling, conventional mechanisms would predict weak or negligible spin pumping. However, as discussed earlier, the linear increase of $g^{\uparrow\downarrow}_{\rm eff}$ with HSQ thickness suggests a bulk spin-sink behavior and a high spin diffusion length. 

One possible explanation for the unexpectedly large $g^{\uparrow\downarrow}_{\rm eff}$ could be a significant contribution from, or even a dominant role of, orbital pumping~\cite{go2023orbital,pezo2024spin} at the NiFe/HSQ interface. Recent studies have reported orbital pumping and orbital torques with long relaxation lengths in light elements and oxide materials with low spin-orbit coupling when interfaced with ferromagnets, such as Ti~\cite{hayashi2024observation} and CuO$_{x}$~\cite{ding2024mitigation}. Additionally, Si has been predicted~\cite{bernevig2005orbitronics} and more recently confirmed~\cite{matsumoto2025observation} to exhibit a large orbital angular moment, further supporting our argument. 

Hence, we propose that the large effective spin pumping observed in our samples may originate from a \emph{combination} of spin and orbital pumping—i.e., spin-orbital pumping—although further detailed measurements and theoretical analysis are required to fully understand this mechanism. Overall, our results confirm a significant spin-pumping effect and establish HSQ as an efficient non-magnetic spin sink layer. Our study also highlights the potential of inorganic polymers for achieving large spin pumping, opening new opportunities in energy harvesting from photovoltaic polymers~\cite{li2022sunlight,guo2024emerging} and and in the development of flexible spintronic devices for memory and wearable healthcare applications~\cite{pachat2022magneto}.

\section{Conclusion}
In summary, we investigated the magnetodynamic response of HSQ/NiFe thin films and demonstrated significant spin pumping from NiFe into the HSQ layer, as evidenced by linewidth broadening and increased effective Gilbert damping. Our findings provide strong experimental support for efficient spin injection into HSQ, a silicon-based inorganic polymer. The spin injection efficiency, quantified by the effective spin mixing conductance, was determined to be $19.8~{\rm nm^{-2}}$ for a 48 nm thick HSQ thin film, which is remarkably high and comparable to values reported for heavy metals with strong spin-orbit coupling. Given that HSQ is an insulating material with inherently low spin-orbit coupling, our results suggest that orbital pumping may play a dominant role in spin transport at the NiFe/HSQ interface. Further theoretical and experimental investigations are necessary to fully elucidate the contribution of orbital effects in HSQ-based spintronic systems. Beyond its fundamental implications, our study establishes HSQ as a promising non-magnetic spin sink for next-generation spintronic applications. The possibility of achieving efficient spin transport in an inorganic polymer opens new avenues for flexible and energy-efficient spintronic devices. Future research could explore the integration of HSQ-based heterostructures into energy harvesting technologies and wearable electronics.

\section*{Methods}
\subsection*{Sample Preparation} The bilayer thin films of HSQ/NiFe were prepared using a combination of spin coating and DC/RF magnetron sputtering (See Fig. 1a-e). First, the polymer hydrogen silsesquioxane (HSQ, $\rm H_8Si_8O_{12}$) was spin-coated onto high-resistivity silicon (HiR-Si) substrates. We deposit these polymer thin films using 2\% HSQ resin solution in methyl isobutyl ketone (MIBK) solution, the low viscosity of the solution gives higher uniformity. No pre or post-annealing treatment were performed. The HSQ solution (2\% HSQ in MIBK) was spin-coated under optimized conditions to achieve film thicknesses of 30, 36, and 48 nm. Following this, NiFe thin films with varying thicknesses (4, 5, 7, and 10 nm) were deposited using DC/RF magnetron sputtering. The sputtering was conducted in an ultra-high vacuum chamber (AJA Orion 8) with a base pressure of $4 \times 10^{-8}$ mbar. The deposition of NiFe thin films was carried out with a DC power of 50 W and a low deposition rate of 1.4 Å/s to avoid potential damage to the HSQ layer. All films were capped with a 3 nm Al$_2$O$_3$ layer to prevent oxidation.

Four sets of samples were prepared for spin-pumping analysis: (i) a reference set of NiFe thin films without HSQ, (ii) HiR-Si/HSQ/NiFe bilayers, (iii) a reversed stack where NiFe was deposited first, followed by immediate HSQ spin-coating to minimize oxidation, and (iv) a control set of HSQ-coated Si substrates with no NiFe layer. The varying thickness HSQ-coated substrates, along with bare Si substrates, were placed in the sputtering chamber together to ensure consistent NiFe/Al$_2$O$_3$ thin films. All samples were fabricated under identical deposition rates to ensure consistency across the series. Surface characterization, presented in Fig. 1f, was performed using atomic force microscopy (AFM) in tapping mode with a Bruker Dimension Icon system.

\subsection*{Atomic Force Microscopy} Atomic force microscopy (AFM) has been done using high resolution Bruker Dimension ICON system to examine the surface morphology in tapping mode under ambient conditions. The scans were conducted over an area of $5\times5\rm {\mu m^2}$ and $1\times1\rm {\mu m^2}$with a pixel size of $512\times512$ pixels. The obtained AFM images were analyzed using Nanoscope analysis software to obtain the roughness values. The root mean square roughness was found from the range of 0.12-0.36 nm, which confirms uniformity and defect free nature of the thin films which ultimately supports the high quality of the stacks.

\subsection*{Ferromagnetic Resonance} Ferromagnetic resonance (FMR) measurements were performed to investigate the magnetodynamical properties using a coplanar waveguide (CPW)-based NanOsc PhaseFMR-40 setup. The FMR measurements were carried out at room temperature over a frequency range of 3–15 GHz with an in-plane external magnetic field. The sample was positioned downward (flip-chip), i.e., with the deposited layers facing the CPW. An electromagnet was used to produce an external magnetic field ($\mu_0 H$) of up to 1 T. A small frequency (98 Hz) field modulation was employed with the Helmholtz coil (biased with an integrated Helmholtz coil source in the instrument) to boost the signal-to-noise ratio. All measurements were performed at ambient temperature to ensure reproducibility.

\textbf{Supporting Information} \par

Supporting Information file is available from the corresponding author.

\section*{Acknowledgment}
S.G. and H.S. acknowledge the Central Research Facility at The NorthCap University, Gurugram for its instrumental and financial support. J.Å. acknowledges funding from the Knut and Alice Wallenberg Foundation (Dnr.~KAW 2022.0079) and as a Swedish Research Council Distinguished Professor (Dnr.~2024-01943).

\textbf{Conﬂict of Interest} \par
The authors declare no conflict of interest. 

\textbf{Author Contributions} \par
  S.G., A.K., A.R. and J.Å. initiated the idea and planned the study. S.G. prepared the thin films, obtained experimental data and performed the FMR analysis with input from all authors. H.B. performed the AFM measurements. S.G. wrote the original draft of the paper. A.K., A.R. and J.Å. coordinated and supervised the work. All authors contributed to the data analysis and co-wrote the manuscript.\\

\bibliography{reference}

\end{document}